\documentclass[10pt,a4paper]{article}
\usepackage{epsf,amsmath,amssymb}
\usepackage[dvips]{graphicx}

\begin{document}

\begin{center}
 \textbf{
 Similarity and Dissimilarity between Influences of Anchoring Walls and
 of External Fields on Nematic and Smectic A Phases
 }
\end{center}
\begin{center}
 M. Torikai and M. Yamashita\\
 Department of Physics Engineering,
 Faculty of Engineering,
 Mie University,
 Kamihama 1515, Tsu, Mie Prefecture, Japan, 514-8507
\end{center}

\begin{quote}
 The McMillan liquid crystalline models under the influence of
 homeotropic anchoring walls and of external fields are investigated.
 For thin systems, the existence of the critical thickness, below which
 the system does not undergo a discrete phase transition, is confirmed.
 Apparent differences between the influence of the anchoring walls and
 of external fields are elucidated by investigating the order parameters
 and a temperature vs external field phase diagram for the bulk
 systems. 
\end{quote}

\begin{quote}
 \textbf{Keywords:} smectic A phase, nematic phase,
 homeotropic anchoring, external field
\end{quote}

\section{Introduction}
The anchoring conditions at the boundary walls strongly influence the
behavior of the liquid crystals.
The effects of the anchoring is remarkable at isotropic(I)-nematic(N)
phase transition, since the discontinuity of extensive variables at I-N
transition is rather small.
For instance, in a thin liquid crystal system bounded by two parallel
walls at which the liquid crystal molecules are anchored
homeotropically, the I-N transition occurs at higher temperature than
the transition temperature of the bulk
system~\cite{ShengPRL1976,Yokoyama1988,Sluckin1990,Poniewierski1987}.
Furthermore if the homeotropic anchoring is strong and the system is
thin enough, the I-N transition disappears and the system undergoes a
continuous change between high- and low-temperature phases.
It is found that there is a critical thickness below which the system
has no transition~\cite{ShengPRL1976,Yokoyama1988}.

The effects of the anchoring at boundaries are similar to the effects of
a uniform external field, as indicated in the early study by
Sheng~\cite{ShengPRL1976}.
Under the external field which aligns liquid crystal molecules, the I-N 
transition temperature shifts higher, as under the influence of
homeotropic anchoring at boundaries.
Corresponding to the existence of the critical thickness for I-N
transition in thin systems, there is a critical strength of the
external field; 
i.e., the I-N transition vanishes and becomes a continuous change under
the external field stronger than the critical external
field~\cite{Wojtowicz,Lelidis1993}.
The similarities between the influence of the external field and of the
anchoring boundary condition were discussed by Poniewierski and 
Sluckin~\cite{Poniewierski1987}. 
In ref.\cite{Poniewierski1987}, they investigated the boundary effects by 
considering a boundary ordering potential at the walls;
they compared the effects of external field and the effects of anchoring
walls by relating the external field and averaged anchoring potential
over the bulk.
The predictions of the averaged anchoring potential approximation are
well satisfied if the anchoring potential is weak.

In the present paper, we study the influence of anchoring walls and
of external fields upon the N and smectic A (A) phases.
Then we discuss the differences between the influence of the anchoring
conditions and of external fields.
We use so-called McMillan's liquid crystal model, which exhibits I, N,
and A phases~\cite{McMillan}.
We consider the situation in which the liquid crystal system is
sandwiched between two parallel walls and the liquid crystal molecules
are strongly anchored homeotropically at the walls.
Changing the temperature and the thickness of the system, we investigate
the temperature dependence of the order parameters and transition
behaviors.
In addition we introduce effective fields due to the inhomogeneity of
the order parameters, and compare the behaviors of the system under the
effective fields and of the system under the external fields.

\section{Formulation}
In this paper, we use the formulation given in ref.\cite{Torikai2004}
for McMillan's liquid crystal model under the external fields and under
the strong anchoring boundary condition, which is carried out in the
framework of the molecular field approximation.

First we introduce the self-consistent equations for a bulk system under 
uniform external fields conjugate to the order parameters.
We denote the N and A order parameters by $s$ and $\sigma$,
respectively. 
These order parameters are defined, as was done in McMillan's original
work, as $s=\langle P_{2}(\cos \theta)\rangle$ and
$\sigma=\langle \cos(2\pi z/d) P_{2}(\cos \theta)\rangle$ where $z$ and
$\theta$ are, respectively,  the $z$-coordinate ($z$-axis is taken to be
parallel to the director) and polar angle of a molecule.
Here we assume that the smectic layer thickness is $d$.
Within the mean field approximation, the one-body potential for McMillan
model is 
\begin{equation}
  V(z, \cos \theta)=
  -V_{0}
  \left[
   s+\alpha \sigma \cos\left(\frac{2\pi z}{d}\right)
  \right]
  P_{2}(\cos \theta), \label{eq:McMillanPotential}
\end{equation}
where the function $P_{2}$ is the second order Legendre function
$P_{2}(x)=(3x^{2}-1)/2$, and the parameter $\alpha$ is a dimensionless
interaction strength for A phase relative to N phase.
Now we introduce two external fields $h_{s}$ and $h_{\sigma}$
corresponding to the order parameters $s$ and $\sigma$, respectively. 
Then the self-consistent equations for order parameters are
\begin{subequations}
 \begin{align}
  s=
  I(
  \beta(h_{s}+ V_{0} s),
  \beta(h_{\sigma}+ V_{0} \alpha \sigma)),
  \label{eq:sWithField}\\
  \sigma=
  J(
  \beta(h_{s}+ V_{0} s),
  \beta(h_{\sigma}+ V_{0} \alpha \sigma)),
  \label{eq:sigmaWithField}
 \end{align} \label{eq:ssEqWithField}
\end{subequations}
where $\beta$ denotes the inverse temperature.
The functions $I(\eta, \zeta)$ and
$J(\eta, \zeta)$ are
defined as 
\begin{subequations}
 \begin{align}
  I(\eta, \zeta)
  &=
  \frac{\partial}{\partial \eta}
  \ln Z(\eta, \zeta),
  \\
  J(\eta, \zeta)
  &=
  \frac{\partial}{\partial \zeta}
  \ln Z(\eta, \zeta),
 \end{align}\label{eq:IandJ}
\end{subequations}
where $Z$ is
\begin{equation}
 Z(\eta, \zeta)=
  \int_{0}^{\pi} d\theta \sin \theta \int_{0}^{d} dz
  \exp\left\{
       \left[
	\eta+\zeta \cos\left(\frac{2\pi z}{d}\right)
	\right]
       P_{2}(\cos \theta)
      \right\}. \label{eq:partitionFunction}
\end{equation}
For a given set of external fields, the self-consistent equations
\eqref{eq:ssEqWithField} have some sets of solutions corresponding to I,
N, and A phases. 
Among these sets of solutions, the thermodynamically stable set gives
the smallest value of a function
\begin{multline}
 \beta
 F(\beta, h_{s}, h_{\sigma};s,\sigma)
 =\\ 
 \frac{\beta V_{0}}{2} s^{2}
 +\frac{\beta V_{0}}{2} \alpha \sigma^{2}
 -\ln \frac{Z(\beta(h_{s}+ V_{0} s),
 \beta (h_{\sigma}+ V_{0} \alpha \sigma))}
 {Z(0, 0)},\label{eq:FforBulk}
\end{multline}
and the smallest value of $F$ is the thermodynamic free energy. 
The other sets of solutions correspond to the metastable and unstable
states. 

In order to treat the inhomogeneous systems, we use a discrete McMillan
model in which the systems are divided into layers by planes parallel to
the boundary walls.
We choose the layer thickness thin enough to neglect the spacial
variance of order parameters inside the layer;
then we can assume the $n$th layer has its own N and A order parameters
$s_{n}$ and $\sigma_{n}$, respectively.
We assume that this layer thickness and the smectic layer thickness $d$
are of the same order.
In the following we restrict ourselves to the systems which have
thickness $D=(N+2)d$ with integer $N$ (0th and $N+1$th layers are in
contact with walls).
Under these assumptions, the molecular field contributions appearing in
the right-hand side of eqs.~\eqref{eq:ssEqWithField}, $V_{0}s$ and
$V_{0}\alpha\sigma$, are divided into the intralayer and interlayer
molecular field contributions.
Then the self-consistent equations of homogeneous systems is generalized
to the self-consistent equations of inhomogeneous systems as
\begin{subequations}
  \begin{align}
   s_{n}= 
   I(
   \beta(\tilde{h}_{s}^{(n)}+ V_{0} s_{n}),
   \beta(\tilde{h}_{\sigma}^{(n)}+ V_{0}\alpha \sigma_{n})),
   \label{eq:scEq2-1}\\
   \sigma_{n}=
   J(
   \beta(\tilde{h}_{s}^{(n)}+ V_{0} s_{n}),
   \beta(\tilde{h}_{\sigma}^{(n)}+ V_{0} \alpha \sigma_{n})),
   \label{eq:scEq2-2}
  \end{align}\label{eq:scEq2}
\end{subequations}
where $V_{0}=V_{0}' + 2 V_{0}''$;
the parameters $V_{0}'$ and $V_{0}''$ are the intralayer and 
interlayer interaction potentials, respectively.
The quantities $\tilde{h}_{s}^{(n)}$ and $\tilde{h}_{\sigma}^{(n)}$ are
defined as
\begin{subequations}
 \begin{align}
  \tilde{h}_{s}^{(n)}=
  V_{0}''(s_{n-1}-2s_{n}+s_{n+1}),
  \label{eq:effectiveFields1}\\
  \tilde{h}_{\sigma}^{(n)}=
  V_{0}''(\sigma_{n-1}-2\sigma_{n}+\sigma_{n+1}),
  \label{eq:effectiveFields2}
 \end{align}\label{eq:effectiveFields}
\end{subequations}
and these quantities vanish if the system is homogeneous.
We shall call  $\tilde{h}_{s}^{(n)}$ and $\tilde{h}_{\sigma}^{(n)}$ the
\textit{effective fields} because of the similarity between
\eqref{eq:ssEqWithField} and \eqref{eq:scEq2}.
Corresponding to the eq.\eqref{eq:FforBulk}, the thermodynamically
stable set of solutions of eq.\eqref{eq:scEq2} gives the smallest value
of a function 
\begin{multline}
 \beta
 F(\beta, \{s_{n}\},\{\sigma_{n}\})
 =\\
 \sum_{n=1}^{N}
 \left\{
 \frac{\beta V_{0}}{2} s_{n}{}^{2}
 +\frac{\beta V_{0}}{2} \alpha \sigma_{n}{}^{2}
 -\ln \frac{Z(\beta(\tilde{h}_{s}^{(n)} + V_{0} s_{n}),
 \beta(\tilde{h}_{\sigma}^{(n)} + V_{0} \alpha \sigma_{n}))}
 {Z(0, 0)}
 \right\}  \\
 + \sum_{n=1}^{N-1}
 \left\{
 V_{0}'' s_{n} s_{n+1} +
 \alpha V_{0}''  \sigma_{n} \sigma_{n+1}
 \right\}
 ,\label{eq:FforThin}
\end{multline}
where the sets $\{s_{n}\}$ and $\{\sigma_{n}\}$ are the solutions to
eqs.~\eqref{eq:scEq2}. 

We impose strong homeotropic anchoring condition, i.e., the perfect
nematic order is induced by the walls and thus $s_{0}=s_{N+1}=1$.
Furthermore we assume that the boundary walls are smooth planes and thus
the smectic order parameter is also unity at each wall, i.e.,
$\sigma_{0}=\sigma_{N+1}=1$. 

\section{Results and Discussions}
The McMillan model exhibits qualitatively different phase behavior
depending on the value of parameter $\alpha$.
The bulk system undergoes a first-order transition directly from I phase
to A phase for $\alpha \gtrsim 0.98$.
For $\alpha \lesssim 0.98$, N phase appears as an intermediate phase
between the I phase and A phase.
Both the I-N and N-A transitions are first-order in
$0.70 \lesssim \alpha \lesssim 0.98$.
The N-A transition becomes a second-order transition for
$\alpha \lesssim 0.70$ while I-N transition remains a first-order
transition. 
Although it is important to investigate systems of several $\alpha$, we
restrict ourselves here to the system which has three (I, N and A)
phases and the two transitions between them are both first-order.
We choose the parameter $\alpha=0.88$.
The results for the other $\alpha$ values will be published elsewhere.

\begin{figure}
 \begin{center}
  \includegraphics[width=8.5cm]{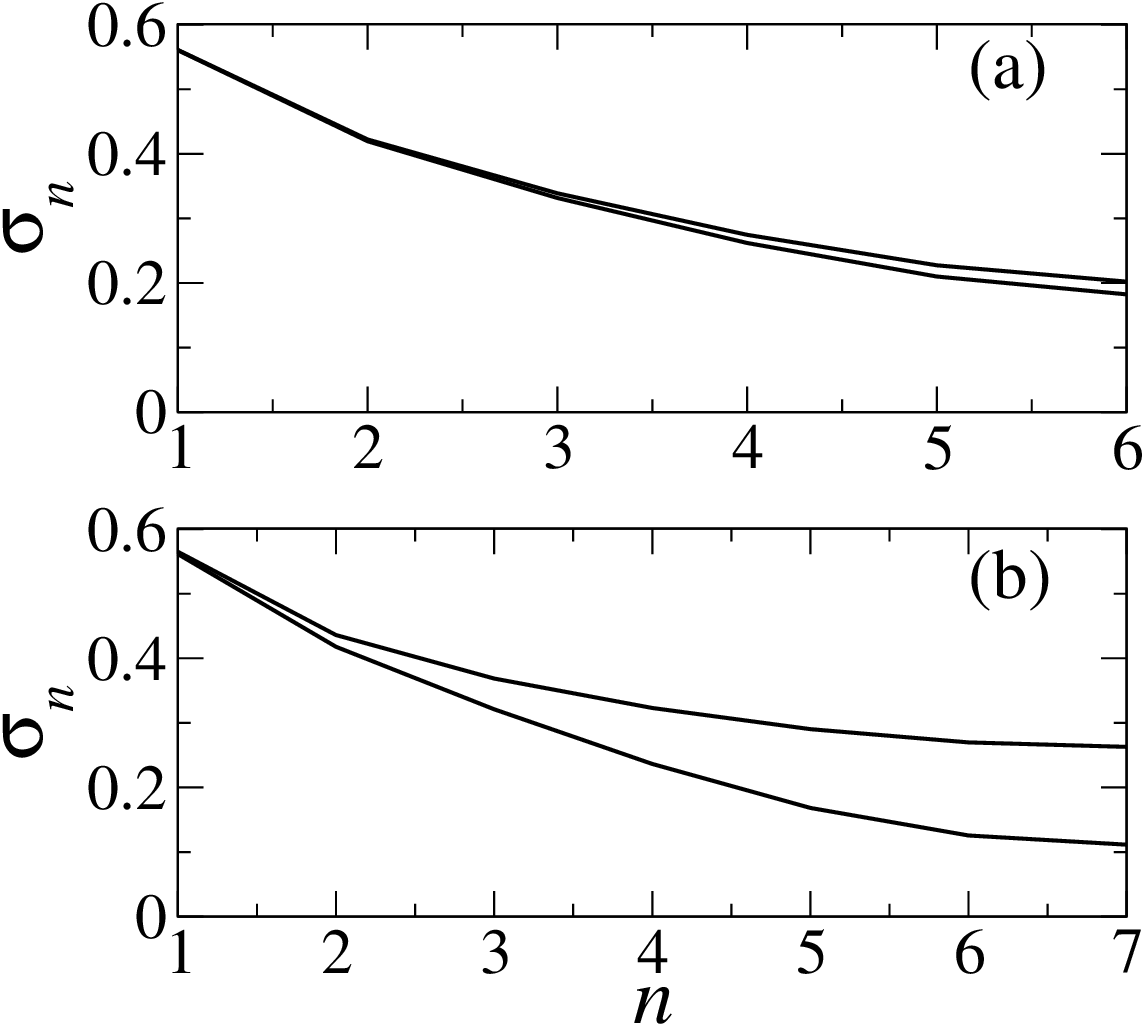}
 \end{center}
 \caption{
 Smectic A order parameter profile $\sigma_{n}$ for the systems with
 thickness (a) $N=12$ and (b) $N=13$.
 In both figures, curves are calculated at temperatures where the order
 parameter difference $\sigma_{n}(T+\Delta T) - \sigma_{n}(T)$ is
 maximum ($\Delta T = 10^{-6}$);
 (a)$T=0.212192$ and $T=0.212193$; (b)$T=0.211898$ and $T=0.211899$.
 } 
 \label{fig:sigmaProfile}
\end{figure}

\begin{figure}
 \begin{center}
  \includegraphics[width=8.5cm]{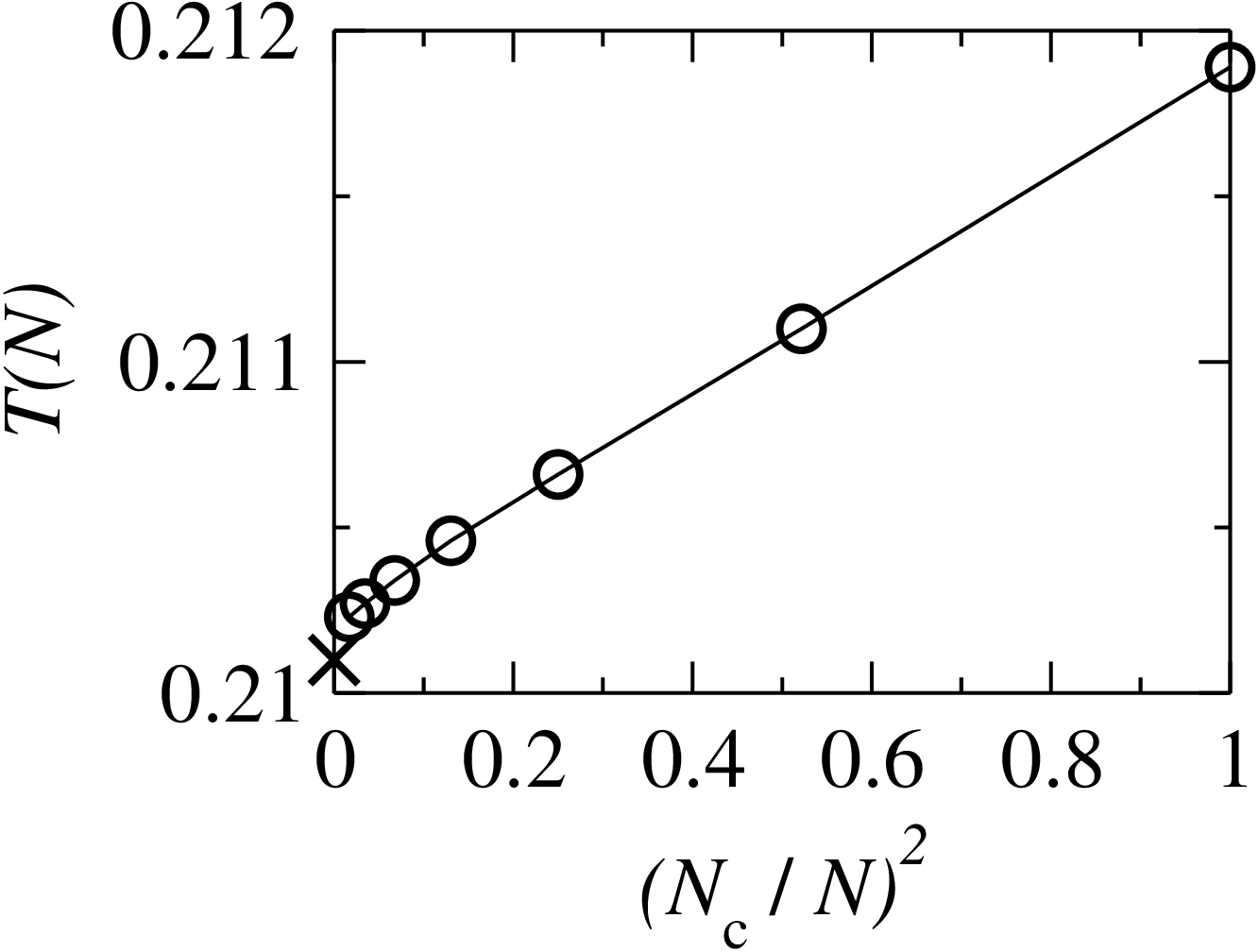}
 \end{center}
 \caption{
 The thickness dependence of the transition temperature $T(N)$ (circles)
 for the system with parameter $\alpha=0.88$.
 The horizontal axis represents the square of inverse thickness
 normalized by the critical thickness $N_{\text{c}}=13$.
 The transition temperature for the bulk system is indicated with a
 symbol ``$\times$''.
 } 
 \label{fig:Ninv2vsT}
\end{figure}

We show in fig.~\ref{fig:sigmaProfile} the order parameter profiles for
systems with different thickness.
These profiles are calculated at temperatures near the bulk transition
temperature.
Since fig.\ref{fig:sigmaProfile}(b) exhibits the first order transition
and fig.\ref{fig:sigmaProfile}(a) does not, the critical thickness for
$\alpha=0.88$ is $N=13$.
The transition temperatures of the thin system is slightly higher than
the bulk transition temperature.
The thickness dependence of the transition temperature are shown in
fig~\ref{fig:Ninv2vsT}. 
The change in transition temperature $\Delta T = T(N) - T(\infty)$
behaves as $\Delta T \sim N^{-2}$ for small $N$.
We note that the thickness dependence of $\Delta T$ seems inconsistent
with the prediction of Kelvin equation: $\Delta T \sim
N^{-1}$~\cite{Yokoyama1988,Sluckin1990}. 
However, since the Kelvin equation is obtained by neglecting the
thickness dependence of latent heat and of the surface
tension~\cite{Poniewierski1987}, the inconsistency only indicates that 
our studies are restricted to the very thin systems where the Kelvin
equation is not valid.
Since the difference between influence of external fields and of the
anchoring walls is mainly focused in the present study, the discussion
on relation between our results and Kelvin equation will be published
elsewhere. 

\begin{figure}
 \begin{center}
  \includegraphics[width=8.5cm]{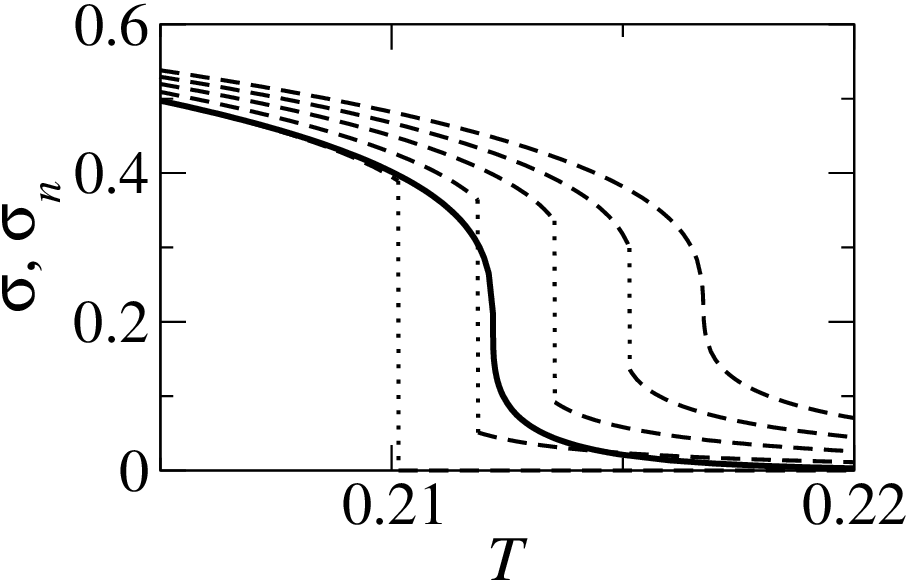}
 \end{center}
 \caption{
 Dashed curves are the order parameters $\sigma(T)$ for bulk systems
 under several external fields: $h_{s}=h_{\sigma}=0$,
 $2.5\times10^{-3}$, $5.0\times10^{-3}$, $7.5\times10^{-3}$,
 and $1.0\times10^{-2}$, from right to left.
 Vertical dotted lines indicate jumps of the $\sigma(T)$.
 The region where the dotted lines appear cannot be reached by
 thermodynamically stable bulk systems. 
 Solid curve is the order parameter
 $\sigma_{6}(T)$ for thin ($N=12$) system.
 }
 \label{fig:TvsOrder}
\end{figure}

We can see the difference between the influence of external fields and
of boundary conditions by investigating the order parameters
quantitatively.
The order parameter $\sigma_{n}$ for $n=6$ of thin
($N=12$) system and order parameter $\sigma$ of bulk system
under some external fields are shown in fig.\ref{fig:TvsOrder}.
There is a region, as shown in fig.\ref{fig:TvsOrder}, where the bulk
system is metastable or unstable no matter how the external fields are.
However an order parameter of a single layer in a thin system pass
through the region where the bulk system is not stable.
In other words, even if we apply the uniform external fields at any
strength, we cannot obtain the same order parameter values of a single
layer. 
This fact clearly shows the difference in the influence of the external
fields and of the effective fields.

In the following, we directly compare a system under uniform external
fields with a single layer under effective fields due to the
inhomogeneity of order parameters, using a procedure originally
introduced in ref.~\cite{Yamashita2003}.
We focus on the system whose thickness is less than the critical
thickness, i.e. the system does not have phase transition.
In order for the direct comparison of external fields with effective
fields, it is better to draw a path of a point
$(\tilde{h}_{s}^{(n)}(T), \tilde{h}_{\sigma}^{(n)}(T) )$
on $T$-$h_{s}$-$h_{\sigma}$ phase diagram of bulk system.
However, since the three-dimensional phase diagram is too complicated,
we consider the $h_{\sigma}$-$T$ phase diagram, not for fixed $h_{s}$
but for $h_{s}=\tilde{h}_{s}^{(n)}(T)$;
i.e., we consider the phase diagram not on the three-dimensional
$T$-$h_{s}$-$h_{\sigma}$ space but on its two-dimensional
subspace defined by the equation $h_{s}=\tilde{h}_{s}^{(n)}(T)$.

\begin{figure}
 \begin{center}
  \includegraphics[width=8.5cm]{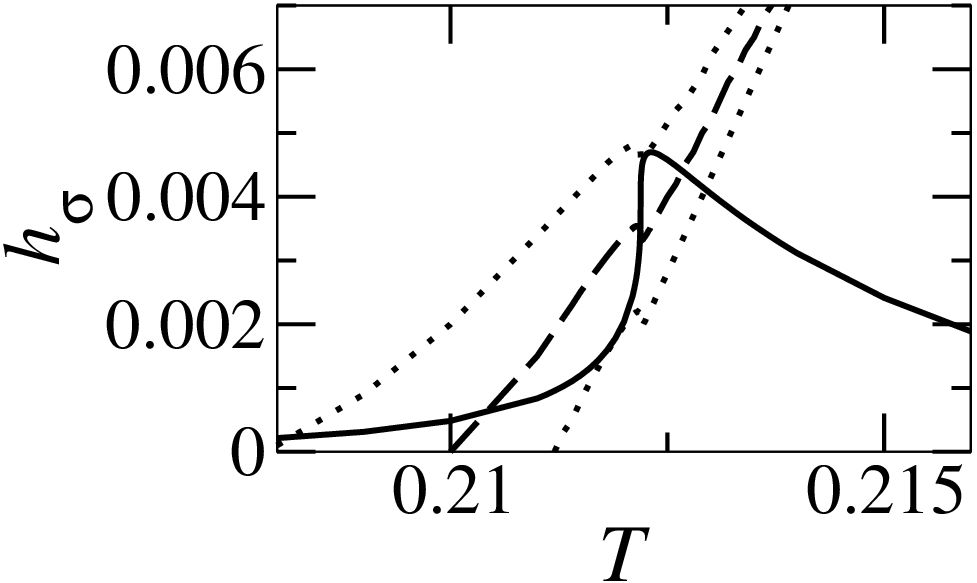}
 \end{center}
 \caption{
 The temperature dependence of effective field $h_{\sigma}^{(n)}(T)$
 ($n=6$ in the system with thickness $N=12$) and the bulk phase
 diagram. 
 Solid curve and dashed curve are the effective field and the coexisting
 curve between N and A phases, respectively.
 Right and left dotted curves are superheating and supercooling curves,
 respectively.
 }\label{fig:hsigmaOnPD}. 
\end{figure}

The $h_{\sigma}$-$T$ phase diagram is shown in fig.~\ref{fig:hsigmaOnPD}. 
In this figure we show, in addition to the coexisting line, the
superheating line and supercooling line.
The N phase is metastable between the coexisting line and supercooling
line, and the A phase is metastable between the coexisting line
and superheating line.
If a bulk system crosses the coexisting line of
fig.~\ref{fig:hsigmaOnPD} from high temperature to low temperature, the
system undergoes a phase transition from N phase to A phase.
The bulk system can change from N phase to A phase without any discrete
transition only if the external field $h_{\sigma}$ is stronger than the
critical value, where the N-A coexisting line terminates.

On this phase diagram, we drawn the effective field
$\tilde{h}_{\sigma}(T)$.
As the temperature decreases, the $\tilde{h}_{\sigma}(T)$ crosses the
coexisting line, touches the supercooling line and the superheating
line, and again crosses the coexisting line.
During this process, the thin system does not undergo any discrete
transition, although the system crosses the coexisting line.
This fact shows that a simple analogy between the effective
field and external field is not valid on discussing the phase
transitions.
Although the thin system is always in stable states, the system seems to
have undergone metastable and unstable states during the process from
high temperature to low temperature.
This is because the stable states of the bulk system and a layer of the
thin system are determined by different functions \eqref{eq:FforBulk}
and \eqref{eq:FforThin}, respectively, though these systems obey the
equivalent equations \eqref{eq:ssEqWithField} and \eqref{eq:scEq2},
respectively.

In summary, we have studied the influence of anchoring walls and
of external fields on N-A phase transition, using a McMillan model.
We have confirmed that the systems sandwiched between homeotropic
anchoring walls exhibit the transition temperature increase and, in
extremely thin systems, disappearance of the phase transition;
these effects of anchoring walls are similar to the effects of external
fields.
However, by investigating the order parameter of a layer and the
effective field, we have found that the influence of the anchoring
walls cannot be understood with a simple analogy of the influence of
the external field.

\end{document}